\documentclass[conference]{IEEEtran}
\IEEEoverridecommandlockouts
\usepackage{cite}
\usepackage{amsmath,amssymb,amsfonts}
\usepackage{algorithm,algorithmic}
\usepackage{graphicx}
\usepackage{textcomp}
\usepackage{hyperref}
\usepackage{booktabs}
\def\BibTeX{{\rm B\kern-.05em{\sc i\kern-.025em b}\kern-.08em
    T\kern-.1667em\lower.7ex\hbox{E}\kern-.125emX}}

\newcommand*\midpoint[1]{\overline{#1}}

\usepackage{lineno,hyperref}
\hypersetup{
     colorlinks   = true,
     linkcolor    = blue,
     citecolor    = red
}
\usepackage{bm}
\usepackage{mathrsfs}
\usepackage{graphicx}
\usepackage{subfigure}
\usepackage[dvipsnames]{xcolor}
\usepackage{nicefrac}
\usepackage{array}
\usepackage{balance}
\usepackage{multirow} 
\usepackage[flushleft]{threeparttable} 

\modulolinenumbers[5]

\makeatletter
\newcommand*{\transpose}{%
  {\mathpalette\@transpose{}}%
}
\newcommand*{\@transpose}[2]{%
  \raisebox{\depth}{$\m@th#1\intercal$}%
}
\newcommand{\thickhline}{%
    \noalign {\ifnum 0=`}\fi \hrule height 1pt
    \futurelet \reserved@a \@xhline
}
\newcolumntype{"}{@{\hskip\tabcolsep\vrule width 1pt\hskip\tabcolsep}}

\makeatother

\ifCLASSINFOpdf

\else

\fi

\begin{document}
\renewcommand{\ttdefault}{cmtt}
\bstctlcite{IEEEexample:BSTcontrol}

\title{Robust Defense Against Extreme Grid Events Using Dual-Policy Reinforcement Learning Agents}

\author{\IEEEauthorblockN{Benjamin M. Peter and Mert Korkali}
\IEEEauthorblockA{\textit{Department of Electrical Engineering and Computer Science}\\ 
\textit{University of Missouri} \\
Columbia, MO 65211 USA \\
e-mail: \{\texttt{bmp792,korkalim\}@missouri.edu}}}

\maketitle

\begin{abstract}
Reinforcement learning (RL) agents are powerful tools for managing power grids. They use large amounts of data to inform their actions and receive rewards or penalties as feedback to learn favorable responses for the system. Once trained, these agents can efficiently make decisions that would be too computationally complex for a human operator. This ability is especially valuable in decarbonizing power networks, where the demand for RL agents is increasing. These agents are well suited to control grid actions since the action space is constantly growing due to uncertainties in renewable generation, microgrid integration, and cybersecurity threats. 
To assess the efficacy of RL agents in response to an adverse grid event, we use the \textit{Grid2Op} platform for agent training. We employ a proximal policy optimization (PPO) algorithm in conjunction with graph neural networks (GNNs). By simulating agents’ responses to grid events, we assess their performance in avoiding grid failure for as long as possible. The performance of an agent is expressed concisely through its reward function, which helps the agent learn the most optimal ways to reconfigure a grid’s topology amidst certain events. To model multi-actor scenarios that threaten modern power networks, particularly those resulting from cyberattacks, we integrate an opponent that acts iteratively against a given agent. This interplay between the RL agent and opponent is utilized in $N-k$ contingency screening, providing a novel alternative to the traditional security assessment. 
\end{abstract}


\section{Introduction}
Power grids are intrinsically complex systems due to the requirement of nonlinear control across massive networks \cite{1216152}.
This complexity grows in modern grids due to the heightened prioritization of their sustainability and security. Integration of renewable generation, microgrids, and cyber attack mitigation features significantly expands the decision space for grid operators, making it challenging to make informed actions efficiently. These decarbonization and security efforts make actions pertaining to the grid’s topology, generation control, and load management incredibly difficult for human operators and motivate the use of reinforcement learning (RL) for quick decision making \cite{ma2024efficient}.

This increase in complexity is readily apparent in grid topology optimization, which motivates substantial efforts to apply RL to this field \cite{rothschild2021deep}. Traditional control methods using certain parameters, including node degree, clustering coefficient, characteristic path length, and betweenness centrality can offer concise insight into topologically complex networks. However, simplification efforts often only offer vague conclusions for decision making \cite{chen2013study} and are unsuitable for extreme grid conditions arising from weather and attacks \cite{liu2022deep}, \cite{nguyen2021smart}. This motivates efforts toward integrating RL algorithms into power grid operation to promote stability during regular and severe grid conditions. Competitions such as ``Learning to Run a Power Network'' (L2RPN) encourage cutting-edge solutions to increasing operational complexity using artificial intelligence and machine learning \cite{marot2021learningrunpowernetwork}. 

The performance of RL in power grid management is often assessed using entities called agents that autonomously take actions based on observations in a given environment \cite{ribeiro2002reinforcement}. Emphasis is often placed on these agents’ performance during the simulation of high-impact, low-probability scenarios such as hurricanes \cite{dehghani2021intelligent}. We aim to expand on this work by applying a novel RL agent configuration to security assessment methods, namely, $N-k$ contingency screening. This screening process can be scaled to exhibit events ranging from routine failures to blackout-inducing catastrophes. An agent model will be assessed across this range of events to demonstrate proficiency in varying conditions. Our main contributions are summarized as follows:
\begin{itemize}
    \item A dual-policy RL agent model integrating GNNs with the PPO algorithm;
    \item Model tuning specifications for maintaining sustainability during extreme grid events; and
    \item Security assessment of this RL agent using $N-k$ contingency screening.\footnote{To our knowledge, this is the first work to analyze RL agents' performance in a contingency screening context, and thus, we propose this proof of concept.}
\end{itemize}

The remainder of this paper is organized as follows. Section~\ref{sec:RL_PG} outlines applications of RL to power systems and discusses related work. Section \ref{sec:framework} provides our RL agent configuration using the PPO algorithm in conjunction with GNNs, wherein we also formulate the agent's reward function, dual-policy model, and opponent integration for security assessment. Section \ref{sec:results} presents simulation results for a case study. Finally, Section \ref{sec:conclusions} provides concluding remarks with respect to future work.

\section{Reinforcement Learning Agents for Power Grid Operation}\label{sec:RL_PG}

\subsection{Related Work}

Previous work has introduced highly specialized agents tuned to mitigate certain adverse grid events. Curriculum-based RL agents are proposed for thermal cascading prevention in \cite{matavalam2022curriculum} and distribution load restoration in \cite{9903581}. Reference\cite{zeng2022resilience} focuses on defense against adversarial cyberattacks using a multi-agent RL model.

This multi-agent RL approach has shown an aptitude for grid operation in a wider variety of scenarios and is readily accessible through frameworks like \textit{PowerGridworld} \cite{biagioni2022powergridworld}. It should be noted that these models differ significantly from dual-policy RL agent models, which will be outlined in depth later in this paper.

Some work has introduced lightweight rule-based models as an alternative to RL agents. Comparably similar performance was demonstrated between an improved rule-based greedy agent and RL models in \cite{lehna2023managing}. Grid power input is optimized in \cite{zhou2023new} using an RL agent in conjunction with a rule-based expert system. For this study, however, we focus explicitly on RL agents due to their theoretically higher flexibility to environments and scenarios.

\subsection{\textit{Grid2Op}}
To configure and simulate RL agents, we use the \textit{Grid2Op} framework \cite{grid2op}, which is built to test sequential decision-making scenarios in power systems. This framework includes modules for tracking time series and actions, emulating certain behaviors, and a backend for power flow computation. The non-linear simulation and RL integration capabilities of \textit{Grid2Op} make it ideal for the objectives of this study. The aforementioned sequential decision process is a Markov decision process (MDP) that interacts with the backend to accurately reflect real-world grid conditions. Note that MDP is commonly used as a heuristic model for strategizing grid operation without RL integration as in \cite{9637871}, but \textit{Grid2Op} facilitates RL models to build upon the MDP mathematical model \cite{qiu1999dynamic}.

Power grids in \textit{Grid2Op} have a graphical structure with nodes and edges analogous to buses and power lines, respectively. Each node and edge has a variety of attributes that may act as parameters in the agent training process. Some examples of these attributes are active and reactive power at nodes and thermal limit and line status at edges. These variables are used by the backend for all computations where they are subject to Kirchhoff’s Laws.

\subsection{Simulation Environment}
\begin{figure*}[htbp]
\centerline{\includegraphics[width=0.85\linewidth]{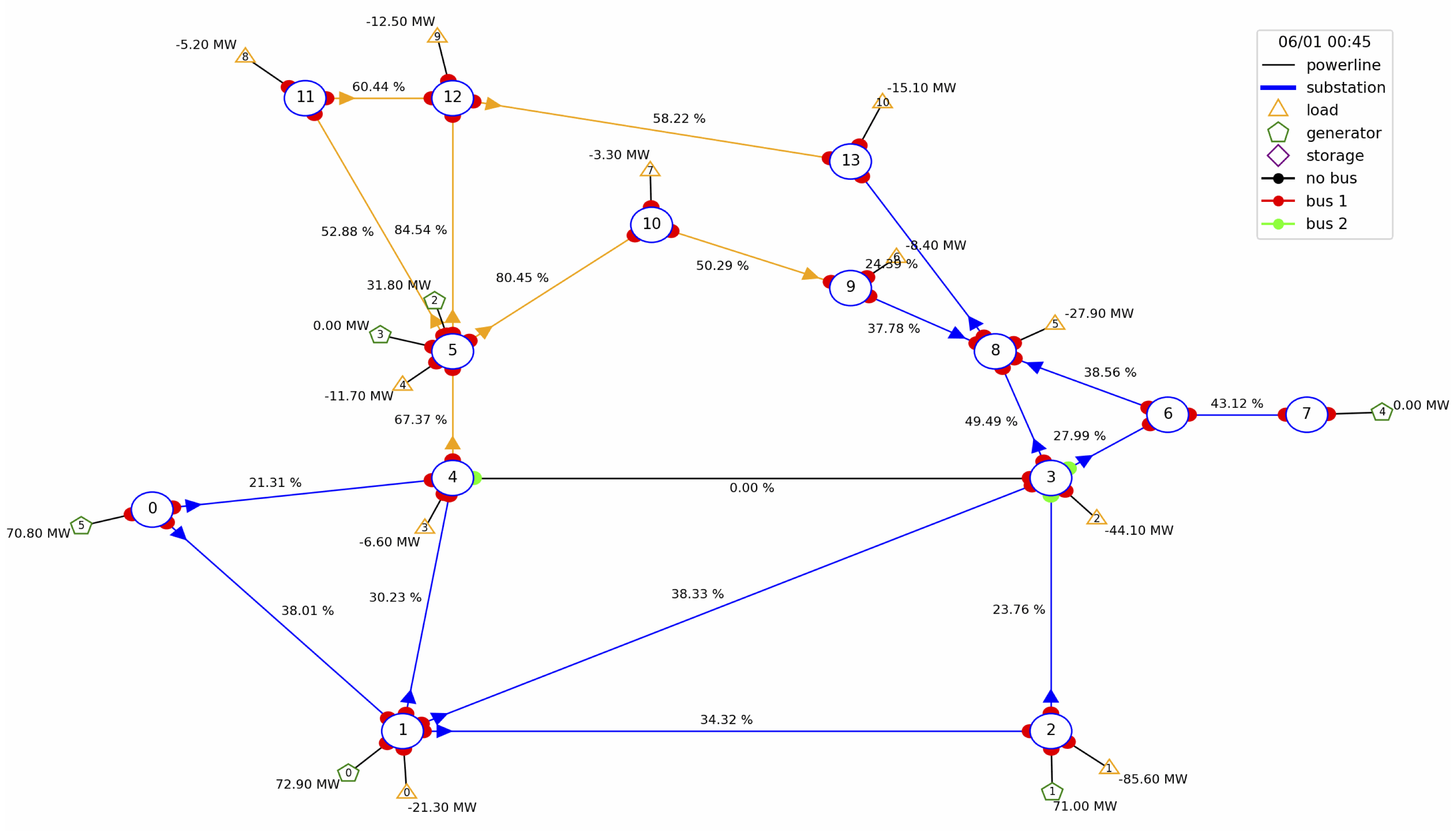}}
\vspace{-.15in}
\caption{Episode visualization for a certain time step on the modified IEEE 14-bus test case.}
\label{fig:episode_viz}
\end{figure*}

Simulation of a given scenario in \textit{Grid2Op} must be run in a grid environment. We use the slightly modified IEEE 14-bus system environment containing $14$ substations, $20$ lines, $6$ generators, and $11$ loads for this proof-of-concept work. This grid is visually represented in Fig.~\ref{fig:episode_viz}. The computationally burdensome nature of the topological action space is already evident on this relatively small test grid for which there are $1.46\times10^{15}$ possible actions \cite{moradi2024heterogeneous}.

An agent interacts with an environment at time step $t$ by deciding on an action that the environment processes and then returns a reward and new state, which is realized by the agent at $t+1$. Since these actions modify the environment topologically, it is important that the environment is explicitly reset between simulations. We prioritize developing an agent that is flexible to a variety of environments while retaining proficiency for maintaining grid stability.

\section{Proposed Framework}\label{sec:framework}
We now discuss the novel RL agent configuration used for efficient grid management in extreme conditions. The grid environment is supported by the \textit{Grid2Op} framework with a \textit{pandapower}\cite{pandapower2018} backend for power flow based on the Newton-Raphson method. This section provides key information on how our custom agent is trained and implemented for proficiency in contingency screening security assessments.

\subsection{Proximal Policy Optimization} \label{sec:ppo}
PPO, developed by OpenAI in 2017 \cite{SchulmanWDRK17}, offers a robust and scalable RL algorithm that can be applied to grid operation agents. As a policy gradient method, PPO is built on the standard equation for the gradient estimator
\begin{equation}
\label{eq:ghat}
\widehat{g}=\mathbb{E}_t[\nabla_\theta\log\pi_\theta(a_t|s_t)\widehat{A}_t].
\end{equation}
Here, $\widehat{A}_t$ estimates the advantage function, and we denote the probability of an Action $a_t$ in State $s_t$ under parameters $\theta$ to be $\pi_\theta(a_t|s_t)$. This $\widehat{g}$ is the differential of the objective function
\begin{equation}
\label{eq:L_theta}
L(\theta)=\mathbb{E}_t[\min(r_t(\theta)\widehat{A}_t,\text{clip}(r_t(\theta),1-\epsilon,1+\epsilon)\widehat{A}_t],
\end{equation}
where $\epsilon$ is the clipping parameter and $r_t(\theta)$ is the probability ratio between policies. While PPO here introduces a clipped surrogate objective, it remains a first-order optimization problem that avoids the complexities of methods like trust-region policy optimization \cite{SchulmanLMJA15} given by
\begin{equation}
\begin{aligned}
\label{eq:1}
& \underset{\theta}{\max} 
& & L(\theta)=\mathbb{E}_t[r_t(\theta)\widehat{A}_t]\\
& \text{subject to}
& & \mathbb{E}_t[-\log(r_t(\theta))]\leq\delta,
\end{aligned}
\end{equation}
where $\delta$ is the maximum allowable Kullback–Leibler (KL) divergence.
The PPO algorithm is implemented in PyTorch for use in \textit{Grid2Op} using a \textit{Stable-Baselines3} \cite{stable-baselines3} library with the \textit{Gymnasium} API\cite{towers2024gymnasiumstandardinterfacereinforcement}.

\subsection{Custom Reward Function}\label{sec:crf}
The PPO policy is adjusted throughout the agent’s training process in response to a custom reward function. Multiple renditions of this function were developed and tested for grid security assessment applications before selecting the one described here. In this case study, we prioritize maximizing grid survival time over other metrics, and the reward system is tuned accordingly. Naturally, it can be altered as necessary for other power grid applications to more highly weighted metrics such as line utilization and economic efficiency.

Despite its intended application to extreme grid events, we determined that a relatively conservative action policy was ideal for lengthened survival time. This reward function also avoids unnecessary complexities and primarily focuses on rewarding no action, penalizing arbitrary or uninformed changes, and avoiding the overload threshold in lines. Here, the action reward {$R_a$} is defined as
\begin{equation}
\label{eq:R_a}
R_a=
\begin{cases}
\gamma \text{ if no action} \\
\delta \text{ if any action} \\
\eta \text{ if minimal action}
\end{cases}
\end{equation}
where positive $\gamma\gg\eta$ and $\delta<0$ are assigned for the given application. Additionally, the rewards scale logarithmically over time so that the agent prioritizes sustained grid survival over short-term rewards. This survival time reward, {$R_s$}, is given by
\begin{equation}
\label{eq:R_s}
R_s=\alpha \log(t+1),
\end{equation}

\noindent for time steps, $t$, and scaling constant, $\alpha$. The reward overload avoidance, $R_o$, is determined by summing the line loading as a ratio of its capacity, $\rho$, across all $N$ lines
\begin{equation}
\label{eq:R_o}
R_o=-\beta \sum_{i=0}^{N}1(\rho_i>\rho_{\text{threshold}}),
\end{equation}

\noindent where $\beta$ is the penalty coefficient.

The total reward, $R_t$, is the sum of these rewards:
\begin{equation}
\label{eq:R_t}
R_t=R_a + R_s + R_o.
\end{equation}
The cumulative reward across the episode of $T$ steps can be given by
\begin{equation}
\label{eq:R_eps}
R_{\text{episode}}=\sum_{t=1}^{T}R_t.
\end{equation}
\subsection{Graph Neural Networks}\label{sec:gcn}
GNNs are gaining widespread attention for applications in power grid planning and operation. In a power systems context, a graph’s nodes represent buses, and edges represent lines connecting buses, both of which contain features pertaining to the environment's observation space. Graph convolutional networks (GCNs) are a subset of GNNs that utilize convolutional operations on graph data using learnable filters. Reference\cite{chen2023topological} uses topological GCNs to plan resilient distribution grid expansions. GNNs are integrated with the PPO algorithm in \cite{lopez2022proximal} to solve optimal power flow efficiently.

Our approach implements a two-layer GCN with processing capabilities to dynamically resize observation vector dimensionality for graph convolution. This processing is facilitated by the PyTorch Geometric library\cite{fey2019pytorch_geometric}. The operation of the underlying GCN is given by
\begin{equation}
\label{eq:H_ell1}
\mathbf{H}^{\ell+1}=\sigma(\widetilde{\mathbf{D}}^{-1/2}\widetilde{\mathbf{A}}\widetilde{\mathbf{D}}^{-1/2}\mathbf{H}^{\ell}\mathbf{W}^{\ell}),
\end{equation}

\noindent where $\ell$ is the layer index; $\mathbf{H}^{\ell}$ is the input feature matrix; $\widetilde{\mathbf{A}}=\mathbf{A}+\mathbf{I}$ is the adjacency matrix including additional self-loops; $\widetilde{\mathbf {D}}$ is the degree matrix of $\widetilde{\mathbf{A}}$; and $\mathbf{W}^{\ell}$ is the weight matrix. This GCN implementation includes an activation function $\sigma$ and a rectifier linear unit (ReLU).

In this proof-of-concept case, the observation data is transformed into a compact $128$-dimensional vector, which the dual-policy networks can easily handle. The GCN’s extraction process gives the PPO-based agent insight into the grid’s structural dependencies and offers significant performance improvements for grid-topological management of catastrophic events.

\subsection{Dual Policy}\label{sec:dual_policy}
Incorporating dual-policy learning is a key component for the agent’s success in security assessment. Different network architectures and hyperparameters are used for a critical and general policy such that the agent can ensure stability during both regular and severe grid conditions. Each policy employs an independent instance of the PPO algorithm to be used in succession to the GCN extraction. The neural network architecture and some notable learning parameters are given in Table~\ref{tab:dp_hyperparams}.
\begin{table}[h!]
    \centering
    \caption{Dual-Policy Hyperparameters}
    \label{tab:sample_separate_headers}
    \begin{tabular}{l lll}
        \toprule
        \textbf{Specification}& \textbf{Parameter}& \textbf{General}& \textbf{Critical}\\
        \midrule
        \multirow{2}{*}{Model Architecture}& Hidden Layer 1& $256$& $512$\\
                                   & Hidden Layer 2& $128$& $256$\\
        \midrule 
        \multirow{3}{*}{PPO Specifications}& Learning Rate& $10^{-4}$& $10^{-3}$\\
                                   & Entropy Coefficient& $0.999$& $0.999$\\
                                   & Gamma& $10^{-3}$& $5\times10^{-4}$\\
        \bottomrule
    \end{tabular}
    \label{tab:dp_hyperparams}
\end{table}
We initially experimented with different reward functions for each policy model, but this only slightly improved performance compared to tuning hyperparameters.\footnote{We plan to further refine the hyperparameters in future work to improve the agent’s performance in larger grid environments, especially with different reward functions.}
In scenario simulation, a dual-policy switch mechanism is used to distinguish which policy is most applicable to current grid conditions. The line loading factor is once again used to determine the critical threshold $\rho_{\text{threshold}}$ at which the critical agent takes action, defined by the switching function 
\begin{equation}
\label{eq:a_t}
\!\!\!\!a_t=
\begin{cases}
\pi_{\text{critical}}(s_t), \text{ if $\max(\rho_1,\rho_2,\dots,\rho_N)>\rho_{\text{threshold}}$} \\
\pi_{\text{general}}(s_t), \text{ otherwise}
\end{cases}
\end{equation}

\noindent where $s$ is the grid state at time step $t$; $\rho_i$ is the load on Line $i$ for $i=1, 2, \dots, N$ for $N$ lines; $\pi$ is the policy used; and $a$ is the action taken. This switching ensures grid stability in both regular and extreme conditions and further prevents the agent from making significant actions unless necessary.

\subsection{Opponent}\label{AA}
We implement an opponent actor into the environment to demonstrate the agent’s proficiency at responding to catastrophic grid events. Depending on the intended application, this opponent can be tuned to represent events like malicious physical and cyber attacks or an evolving natural disaster. In this case, we choose to represent the latter by allowing the opponent to disconnect lines deemed critical. We also limit the opponent to performing only topological grid actions, namely, line disconnections. This constraint aligns with our focus on the agent’s performance in contingency screenings as that assessment is of a similar topological nature.
We formulate an opponent with an adjustable attack interval $\tau_{\text{attack}}$ that acts at times $t$ such that
\begin{equation}
\label{eq:tau_att}
t\mod\tau_{\text{attack}}=0.
\end{equation}
In our case study, we use $\tau_{\text{attack}}=1$ to more closely model a continuously evolving, aggressive event. We define the set of all lines $\mathcal{L}$ and a disconnection operator $\mathcal{D(\mathcal{L})}$ such that
\begin{subequations}
\label{eq:L_DL}
\begin{align}
\mathcal{L} &\subseteq (1,2,\dots,N), \label{eq:L}\\
\mathcal{D(\mathcal{L})} &: \text{Disconnect all lines in }\mathcal{L}. \label{eq:DL}
\end{align}
\end{subequations}
We then define a subset of lines that the opponent can attack, $\mathcal{L_{\text{attack}}}$, which is all lines in our case study. We then define highly loaded lines
\begin{equation}
\label{eq:L_high}
\mathcal{L_{\text{high}}}=\{i\in L_{\text{attack}}:\rho_i\geq\rho_{\text{threshold}}\}.
\end{equation}
This allows our opponent to target these critical lines for each Action $a$ as follows:
\begin{equation}
\label{eq:a_t_opp}
a_t=
\begin{cases}
\mathcal{D}(\mathcal{L}_{\text{high}}), \text{ if } \mathcal{L}_{\text{high}}\neq \varnothing \\
\mathcal{D}(\mathcal{L}_{\text{attack}}), \text{ otherwise}
\end{cases}
\end{equation}
These actions place additional stress upon the agent during the security assessment process and thus motivate the thorough training process described previously.

\subsection{Contingency Screening}\label{sec:cont_screen}
A traditional form of security assessment is $N-k$ contingency screening. This process determines the risk of cascading outages and system failure in networks following the failure of $k$ lines. We uniquely combine this screening process with a reinforcement learning agent to demonstrate its robustness against both routine failures (i.e., combinations of $k$ disconnections for $k\leq2$) or relatively extreme events (i.e., combinations of $k$ disconnections for $k>2$).
We define all possible $k$ combinations of $\mathcal{L}$ as $\mathcal{C}$, i.e., that
\begin{equation}
\label{eq:C}
\mathcal{C}=\{\mathcal{S}\subseteq\mathcal{L}: \ |\mathcal{L}|=k\},
\end{equation}
where $|\mathcal{C}|=\binom{N}{k}$. We then simulate across all sets $\mathcal{S}\in\mathcal{C}$ to determine steps survived $T_s$, cumulative agent reward $R_s$, and cascading failures $F_s(t)$ for each $\mathcal{S}$ to compute respective averages:
\begin{equation}
\label{eq:T_bar}
\midpoint{T}=\frac{1}{|\mathcal{C}|}\sum_{\mathcal{S}\in\mathcal{C}}T_\mathcal{S}
\end{equation}
\begin{equation}
\label{eq:R_bar}
\midpoint{R}=\frac{1}{|\mathcal{C}|}\sum_{\mathcal{S}\in\mathcal{C}}R_\mathcal{S}
\end{equation}
\begin{equation}
\label{eq:F_bar}
\midpoint{F}(t)=\frac{1}{|\mathcal{C}|}\sum_{\mathcal{S}\in\mathcal{C}}F_\mathcal{S}(t)
\end{equation}
Numerical results from this contingency screening process will provide insight into the agent's comparative performance in security assessment simulations.

\section{Numerical Results}\label{sec:results}
This section outlines the numerical results demonstrating the comparative performance of the dual-policy PPO agent against an agentless benchmark, denoted \texttt{NoAgent} henceforth. For this case study, we aim to perform a security assessment of the agent using $N-k$ contingency screening. The environment in which the agent is screened is intentionally hostile to emphasize its enhanced ability to withstand evolving threats and extreme grid conditions. This is achieved by integrating the aforementioned opponent and increasing the system loading by $25\%$.

In this study, we most notably prioritize system survival, i.e., avoiding a complete blackout, as the most significant metric for demonstrating the agent’s success. It should be noted that while cascading outages sometimes occurred in less extreme environments, they were exceedingly rare in this case study, as the opponent and loading conditions would often trigger system failure in a single step. As such, the primary focus of this case study was to assess the critical policy model’s response to potentially catastrophic events and the delegation of less threatening events to the general model.
\vspace{-.1in}
\begin{table}[htbp]
\caption{Average Time Steps Survived (with a $100$-Step Limit)}
\centering
\begin{tabular}{|c|c|c|c|c|c|}
\hline
& \multicolumn{5}{c|}{\textbf{Number of Initial Failures} $\boldsymbol{(k)}$} \\ \hline 
\textbf{Case}& $k=1$ & $k=2$ & $k=3$ & $k=4$  &$k=5$\\
\hline
\texttt{Agent} & $95.10$ & $98.97$ &  $99.05$ & $99.13$  &$99.11$\\ \hline 
\texttt{NoAgent} & $70.20$ & $37.45$ & $15.28$ & $4.56$  &$0.93$\\ \hline\end{tabular}
\label{tab:avg_ts_survived}
\end{table}

\begin{figure}[htbp]
\centerline{\includegraphics[width=1\linewidth]{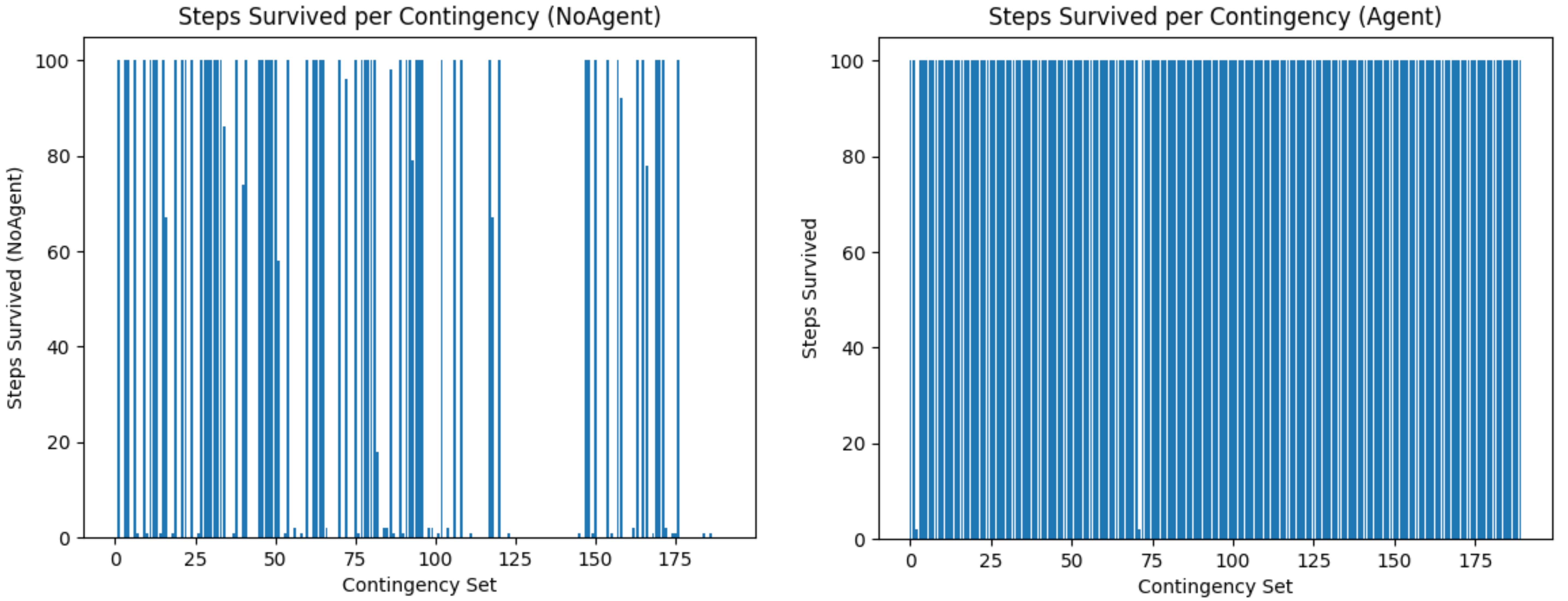}}
\vspace{-.1in}
\caption{Time steps survived for the \texttt{NoAgent} (left) and \texttt{Agent} (right) cases per $N-2$ contingency set.}
\label{fig:ts_N-2}
\end{figure}

\begin{figure}[htbp]
\centerline{\includegraphics[width=1\linewidth]{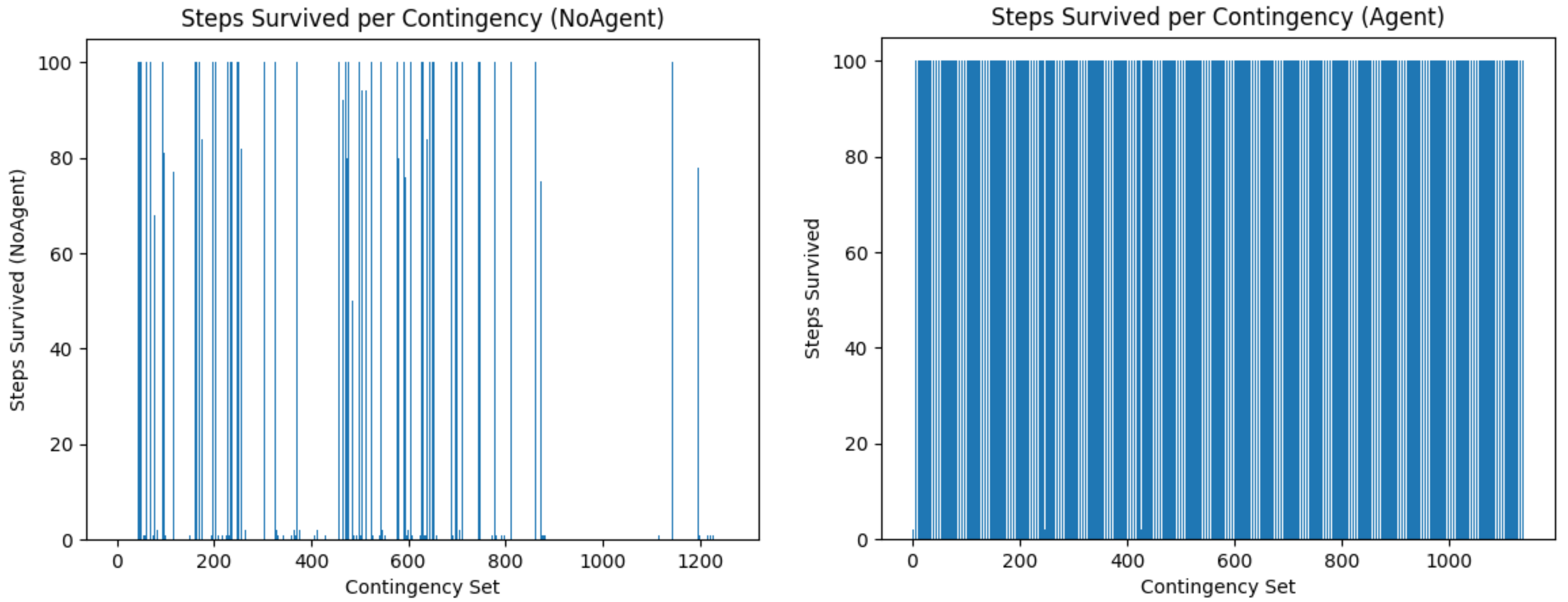}}
\vspace{-.1in}
\caption{Time steps survived for the \texttt{NoAgent} (left) and \texttt{Agent} (right) cases per $N-3$ contingency set.}
\label{fig:ts_N-3}
\end{figure}

The numerical results in Table~\ref{tab:avg_ts_survived} demonstrate the agent’s proficiency in maintaining grid stability amidst extreme grid conditions. Moreover, it is evident that while the \texttt{NoAgent} case performs increasingly worse with heightened $k$ values across all contingency sets, the agent can efficiently adjust for the initial outages and, therefore, offer consistent performance as seen in Figs.~\ref{fig:ts_N-2}
 and \ref{fig:ts_N-3}.

\begin{figure}[htbp]
\centerline{\includegraphics[width=1\linewidth]{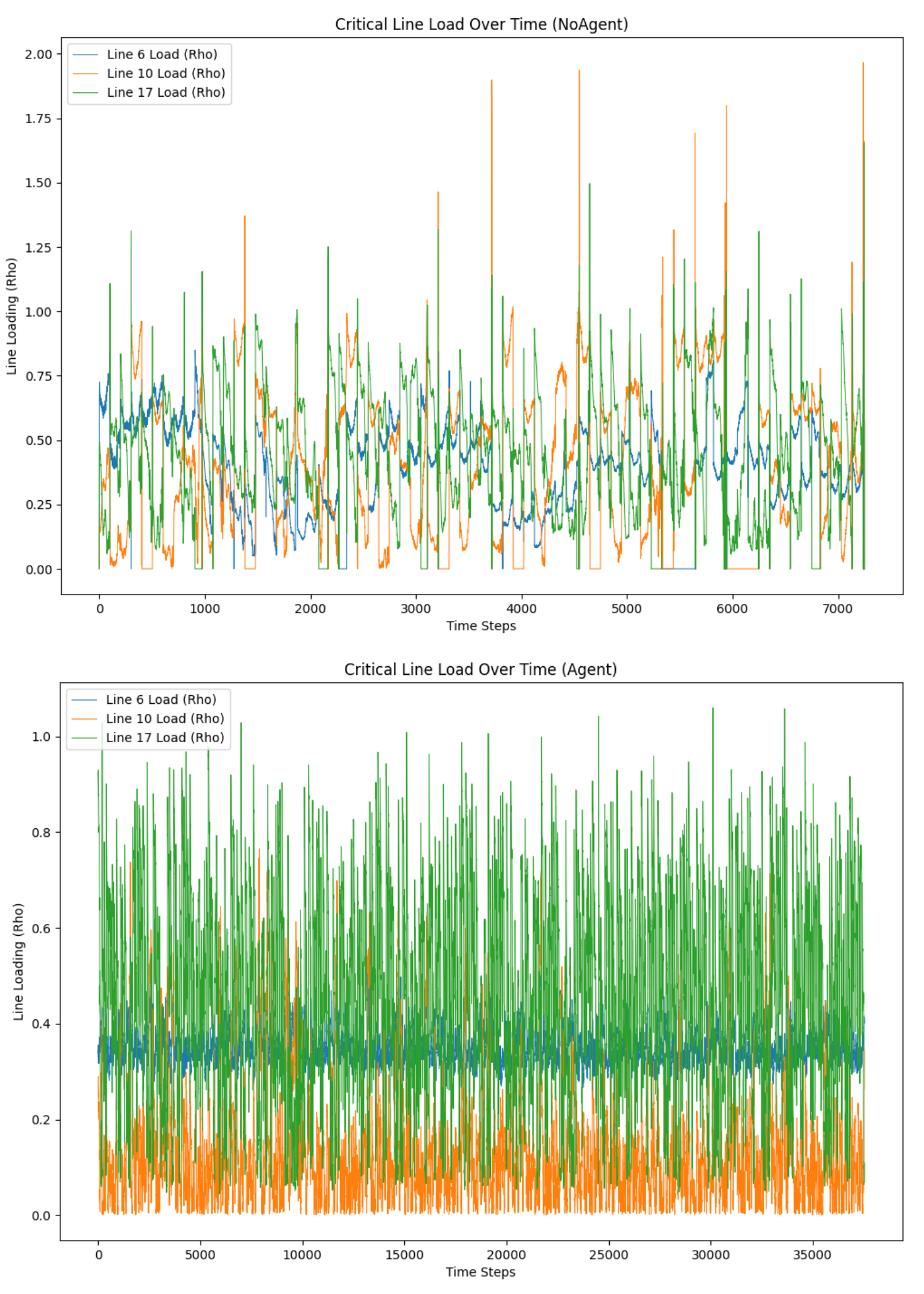}}
\vspace{-.1in}
\caption{The $\rho$ (Rho) values in selected lines across survived time steps for the \texttt{NoAgent} (top) and \texttt{Agent} (bottom) cases, respectively.}
\label{fig:rho}
\end{figure}

\begin{figure}[htbp]
\centerline{\includegraphics[width=1\linewidth]{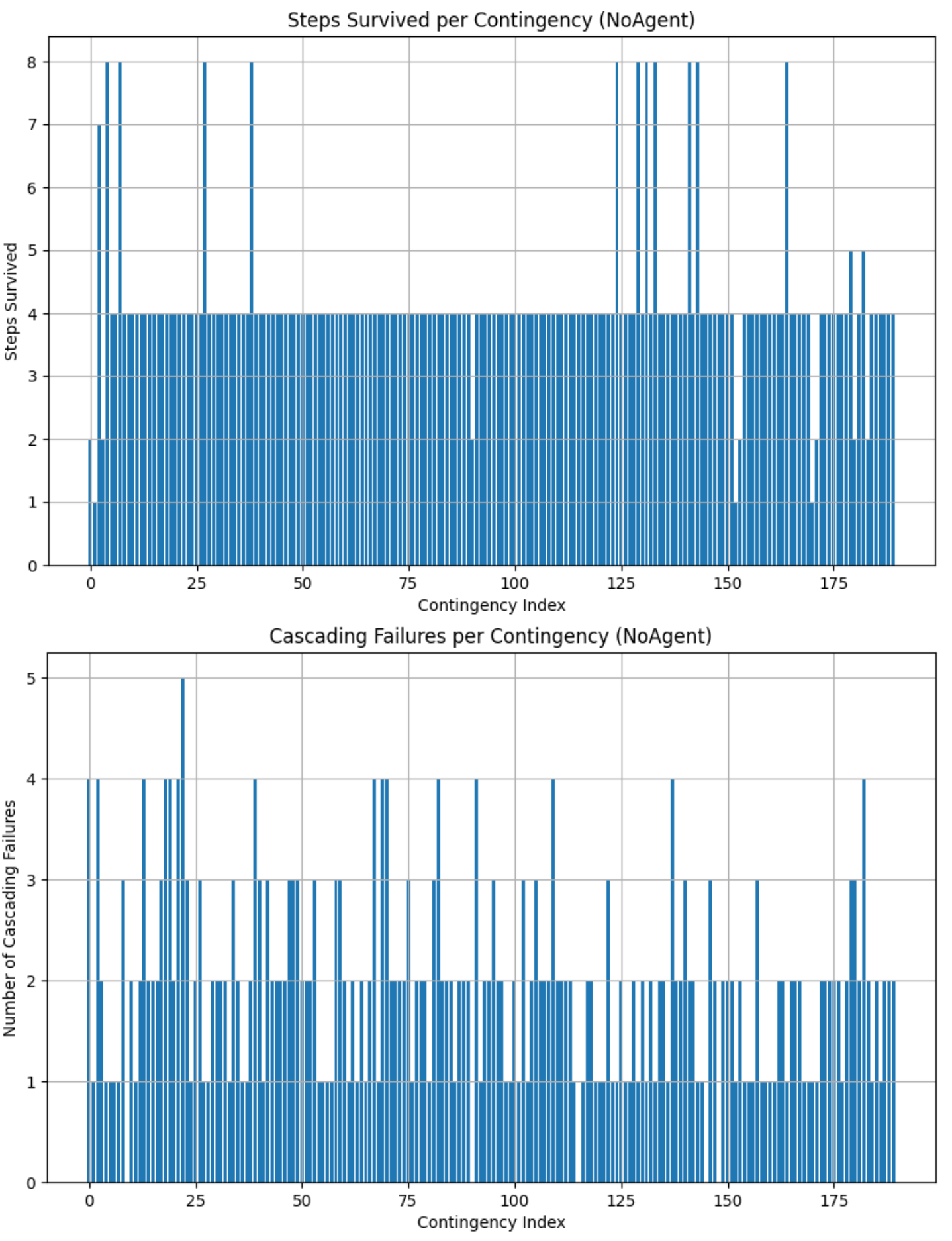}}
\vspace{-.1in}
\caption{Steps survived and cascading failures for an $N-2$ scenario with a modified opponent for the \texttt{NoAgent} case (Note that for the \texttt{Agent} case, the lack of failures makes visualizing cascades irrelevant).}
\label{fig}
\end{figure}

The ratio of line loading, denoted by $\rho$, also provides insight into the agent’s grid-stabilizing capabilities. Maintaining a $\rho$ value less than $1$, i.e., the line’s maximum rated capacity, is essential for avoiding failure-inducing grid conditions. The agent’s critical model is specifically trained to quickly regulate any instances in which this $\rho$ value nears $1$, while the general model prevents abrupt and unnecessary actions at lower $\rho$ values. In the \texttt{NoAgent} case, however, these remedial actions are not possible, and deviations above the maximum loading threshold occur more frequently, as evident in Fig.~\ref{fig:rho}.
When accumulated across all lines, these unstable $\rho$ values quickly lead to system failure. This failure often occurs in a single time step for the \texttt{NoAgent} case when the opponent causes overloading in enough lines simultaneously, leaving the system unsolvable. In systems with modified opponent reward functions, however, cascading failures are often more prevalent for the \texttt{NoAgent} case, as shown in Fig.~\ref{fig}.

\section{Conclusion and Future Work}\label{sec:conclusions}
We successfully developed an RL agent that demonstrates significant proficiency in the security assessment of extreme grid events. This agent integrates a dual-policy model with PPO and GNN for the training process and can be dynamically applied to different power grids. We emphasize the agent's exceptional performance against the opponent for high $k$ values (i.e., $k>2$) as this demonstrates its stabilization capabilities within a severe environment.

A notable observation from agent simulation was a conflict between the objectives of maximizing grid survival time and minimizing cascading failures. This discrepancy was evident when the model was trained to prioritize minimal cascading failures, in which its forecasting would often allow one catastrophic failure to quickly trigger a dead grid state. This prevented failures from slowly accumulating over time but represents a inadequate approach to grid management. As such, we chose to prioritize grid survival, which did in fact greatly reduce cascading failures. However, we emphasize the reconciliation of these conflicting objectives in future work.

Additional future work includes scaling the agent simulation to larger and more realistic power grids to expand upon this proof-of-concept work. There is also potential for further tuning of this agent using intelligent RL hyperparameter optimization.

\ifCLASSOPTIONcaptionsoff
\newpage
\fi

\vspace{-0.2cm}
\bibliographystyle{IEEEtran}
\bibliography{IEEEabrv,References}

\end{document}